\documentstyle[12pt]{article}

\begin{document}

\author{Amy J. Kolan {\dag}, Edmund R. Nowak {\ddag}, and Alexei V. Tkachenko\\
 \it The James Franck Institute, \\
\it The University of Chicago, Chicago, Illinois 60637 \\
{\dag} Permanent Address: \it Physics Department, St. Olaf College,\\
\it  1520 St. Olaf Avenue, Northfield, MN 55057\\
{\ddag} Present Address: \it University of Illinois, Loomis Laboratory of
Physics,\\
\it MC-704 1110 W. Green St., Urbana, IL 61801}
\date{}

\title{\bf On the Glassy Behavior of the Parking Lot Model }

\maketitle

\begin{abstract}

We present a theoretical discussion of the reversible parking problem,
which appears to be one of the simplest systems exhibiting glassy behavior.
The existence of slow relaxation, nontrivial fluctuations, and an
annealing effect can all be understood by recognizing that two different
time scales are present in the problem. One of these scales corresponds to
the fast filling of existing voids, the other is associated with
collective processes that overcome partial ergodicity breaking. The results
of the theory are in a good agreement with simulation data; they  provide a
simple qualitative picture for understanding recent  granular compaction
experiments and other glassy systems.

\end{abstract}

{\bf PACS numbers: 68.45.Da, 61.43.-j, 64.70.Pf}

\newpage

\section{Introduction}

     Constrained dynamics of complex systems has long been  a subject of
extensive experimental and theoretical research.  Certain important
features, such as slowing kinetics,   non-exponential relaxations, and
memory effects are believed to be generic for a wide class of systems
ranging from structural and spin glasses to granular materials and traffic
flows \cite{glass}, \cite{spgl}, \cite{traf}.  In spite of significant
progress in this field there is still no general framework for the
description of jamming and glassy phenomena. The development of a clear
qualitative picture of them has been frustrated by the
relatively high complexity of the considered systems. One could hope to boost the
conceptual progress in this field by analyzing simple models capable of
capturing the important  features of glasses.

 In this paper we present a theoretical discussion of one of the simplest
systems that  exhibits glassy--like relaxation dynamics and a non--trivial
fluctuation spectrum.  This system is  known as the Parking Lot Model (PLM)
\cite{ev}-\cite{eli}, or the continuous Random Adsorption Problem, and it
is defined as follows.  Identical, unit length particles (cars) can adsorb
on a line (curb) at rate $ k_+$ per unit curb length.  They can also leave
the line with rate $k_-$.  The desorption process is unrestricted while the
adsorption is subject to free volume constraints, i.e. two cars cannot
overlap (see Figure 1).  This model can be applied in a straightforward way to
random physical adsorption of large molecules. In addition, the PLM appears
to be one of the most successful models for the description of density
relaxation and fluctuations  in a vibrated granular material. The possible
reason for this is that the dynamics of the PLM drastically depends on the
available free volume, just as in the case of granular materials or
structural glasses.

The dynamics of the  original version of the PLM, in which the  particles
adsorb irreversibly (i.e. $k_-=0$), has been well understood as long ago as
1958 by Renyi \cite{ren}. He  found that the system jams at density
$\rho_c\simeq 0.75$, and the way it approaches this state is given by the
following formula:

\begin{equation}\label{jam}\rho(t)=\int_0^{k_+t}dw\exp\left[-2\int_0^w
du(1-e^{-u})/u\right]
\end{equation}

The late--stage asymptotics of this result is the power law relaxation: $\rho(t)-\rho_c\sim 1/t$.

The desorption process introduced  by  Krapivsky and Ben-Naim \cite{eli},
results in even richer physics. In a recent paper on granular compaction,
Nowak et. al. \cite{ed} presented the results of simulations on the PLM.
These simulations and the experimental granular  compaction data have many
important features in common.  In particular, the average coverage of the
curb  as a function of time, shown in Figure 2, is very similar to  the
density relaxation curve for the  vibrated sand. Once in the steady state,
the finite size of the system results in considerable density fluctuations.
Insight into their dynamics is provided by the power spectrum as seen in
Figure 3.  This figure illuminates one of the most remarkable properties of
the parking lot model, i.e. that it exhibits two very different time scales
at high k values (high density).  These time scales appear in the power
spectrum as two corner frequencies, one at high frequency and one at low.
The low frequency corner is Lorentzian, which indicates that it can be
associated with exponential relaxation at a single time scale.   The high
frequency "corner", however, shows an unusual "hump" that indicates that
the density relaxation cannot be described with only one time scale.
 It is these features in the power spectrum that led Nowak et. al. to
present the parking lot model in conjunction with the sand experiment,
which shows a similar non-trivial fluctuation spectrum.
The existence of several relaxation time scales is a signature  of {\em the
partial ergodicity breaking} exhibited by the  model, i.e.  its
high--frequency evolution does not allow the system to explore all the
configurational space.   In this sense the observed behavior of the PLM may
be  relevant  not only for understanding the particular granular compaction
experiment but also for the whole class of systems exhibiting glassy
relaxation dynamics. Below we focus on developing an analytical description
of the PLM capable of capturing these intriguing features and revealing the
underlying physics.

\section{ The Limitations of the Mean Field Approach}
	
In their original work on the reversible  PLM \cite{eli}, Krapivsky and
Ben-Naim proposed a mean--field description of the problem that can be
essentially  expressed in terms of the following master equation for the
average density, $\rho$:

\begin{equation}\label{mf}\frac{\partial \rho}{\partial
t}=k_+(1-\rho)\exp\left(-\frac{\rho}{1-\rho}\right)-k_-\rho\end{equation}
Here the first term represents the adsorption rate and takes into account
its strong dependence on the available free volume, the second term
corresponds to unrestricted desorption. The above equation indeed captures
some of the features of  the PLM. In particular, it results in an equation
for the equilibrium  density $\rho_{eq}$, that is consistent with the
numerical data:

\begin{equation}\label{eq}\left(\frac{1-\rho_{eq}}{\rho_{eq}}\right)\exp\left(-\frac{\rho_{eq}}{1-\rho_{eq}}\right)=\frac{k_-}{k_+}\end{equation}

Although this mean-field approach does result in typical jamming dynamics,
the predicted density relaxation rate is several orders of magnitude faster
than that observed in the simulations  \cite{ed}.  Moreover, the mean field
description fails to capture the most interesting feature of the PLM, its
fluctuation spectrum. Indeed, the very structure of Eq. (\ref{mf}) cannot
result in anything different from a regular linear response equation for
the density near equilibrium,  i.e. it should have a single relaxation time
for small fluctuations.

An essential feature that the above mean--field approach overlooks is the
strong  correlation between adsorption and desorption events.  This
correlation is familiar to anyone who has ever attempted to park in a big
city.  Cars do not leave the curb very often but as soon as they do an
incoming car rushes in to take the newly created space.  This paired
adsorption-desorption process does not change the density of the system;
it is equivalent to merely sliding a car in its parking space.  One can explicitly emphasize  this by the introduction of a slow
(adiabatic) variable $Z\equiv N+N^*$, which is the sum of the number of
adsorbed cars, $N$, and the number of voids large enough to  fit at least
one  particle, $N^*$.  If the system is sufficiently dense, most individual
adsorption or desorption events do not change the parameter $Z$.   The
existence of this slow variable results in the observed separation of the
relaxation time scales.  The fast modes of
 the density relaxation correspond to the evolution with nearly constant
$Z$, while the slow dynamics is determined by  low-probability  events,
which result in a drift of $Z$. In other words, on short time scales the
ergodicity of the PLM is broken: the system can explore only the part of
the configurational space corresponding to a  constant $Z$.

The fast and slow modes are expressed as two typical time scales in the
density fluctuations and also lead to a two-stage relaxation process of the
density evolution from $\rho=0$ to $\rho_{eq}$. In the case of a very small
desorption rate $k_-$ the system follows the universal
irreversible--adsorption curve, Eq. (\ref{jam}) until it jams at
$\rho_c\simeq 0.75$. Afterward, it slowly evolves towards the equilibrium
density. By using the above result for $\rho_{eq}$ as a function  of
$k\equiv k_+/k_-$ and knowing $\rho_c$ one can construct the "kinetic phase
diagram" of the system (see Figure 4): if $\rho_{eq}(k)<\rho_c$ the system
reaches the  equilibrium density before it jams, otherwise  it relaxes to
$\rho_c$ and then slowly creeps toward equilibrium. It would be tempting to
associate the  critical point $k_c\simeq 60$, which separates the two
regimes, with a glass transition.  However, this term is traditionally
reserved for the hypothetical point at which the kinetic coefficients of a
system would go to zero (in our simple  case this is the point $k=\infty$).
Thus, the zone of partial ergodicity breaking appearing at $k>k_c$ is an
analog of what is conventionally called a supercooled liquid rather than a
glass.

An important implication of the above picture of the PLM kinetics is the
existence of an annealing effect, which is  typical for classical  glassy
systems and has also been observed in the granular compaction experiments 
\cite{ed}.
The idea is that one can overcome the slow kinetics  of the supercooled
system by "heating" it (by decreasing $k$ below $k_c$), then slowly cooling
it (increasing $k$),  so that the system would follow the reversible
equilibrium curve $\rho_{eq}(k)$.

\section{The Fast Dynamics of the System}

     As we have pointed out above, the fast dynamics of the PLM is
dominated by a two part process: a desorption that takes place at rate
$k_-$ leaving a void and an adsorption that occurs at rate $k_+z$, where z
is the void size minus a car length.  This two-part process is equivalent
to simple replacement of one adsorbed particle with another or a sliding of
a single particle in its own space.  As a result, the separations between a
given particle and its two nearest neighbors, $x_1$ and $x_2$, change
randomly to the new values $x_1'$ and $x_2'$ in such a way that their sum
remains constant: $x_1+x_2= x_1'+ x_2'=z$.    This sliding process creates
the necessary mixing to ensure that the steady-state distribution of the
spaces between cars, p(x), corresponds to the maximal entropy for a fixed
sum of all the separations.  Thus, the distribution function   can be
obtained by maximizing the functional
\begin{equation}\Phi[p]=-\int_0^\infty dx\left(p(x)\log
p(x)+p(x)\frac{x}{\delta}\right) \end{equation}
 with respect to $p(x)$, where $1/\delta$ is the  Lagrange multiplier
conjugate to $x$.  Naturally, the resulting formula for $p(x)$ is similar
to the Boltzmann distribution:
\begin{equation}
p(x)=\frac{1}{\delta}\exp\left(-\frac{x}{\delta}\right)
\end{equation}
The parameter  $\delta$  is equal to the average spacing between cars, i.e.
\begin{equation}\delta=\frac{1-\rho}{\rho}\end{equation}
     We now proceed with the calculation of the high frequency part of the
fluctuation  spectrum. The coupled adsorption--desorption process can be
viewed as a relaxation-excitation of the conventional  two-state
(telegraph) system, which  can switch from one state,  '0',   to another
one,  '1',  with a characteristic time $t$ and then relax back to '0' with
the characteristic time $\tau$. The power spectrum of telegraph noise has
been investigated by Machlup \cite{mach}  and is given by
\begin{equation}\label{lor}S(\omega)=\frac{1}{\pi(t+\tau)}\frac{1}{\omega^2+
\nu^2}\end{equation}
where $\nu=1/\tau+1/t$.

The fast dynamics of the density is the superposition of these nearly
independent telegraph  modes, whose number is essentially the number of
cars (or, more precisely, $Z$).
     An important feature of the adsorption--desorption modes is that while
their "excitation rate", $ 1/t=k_-$ is uniform over the system, the
relaxation time is the property of an individual excited state, i.e. the
adsorption rate is proportional to the free volume available at a given
void: $1/\tau=k_+(x_1+x_2)$. As a result, the fluctuation spectrum deviates
from the simple Lorentzian form, Eq (\ref{lor}):

\begin{equation}\label{hi}S_H(\omega)\equiv \langle
\rho_\omega\rho_{-\omega}\rangle\equiv\frac{\langle N_\omega
N_{-\omega}\rangle}{L^2}= \frac{\rho k_-}{\pi
L}\int_0^\infty\frac{f(\nu)d\nu}{\omega^2+\nu^2}\end{equation}
Here $L$ is the total length of the system. Note that here and below we
assume that $k\equiv k_+/k_-\gg 1$, so that the typical adsorption process
is much faster than the desorption. As a result, the desorption rate
determines the amplitude in the above expression, while  the (non--uniform)
adsorption dominates its  frequency dependence (i.e. $\nu\simeq 1/\tau =
k_+ (x_1+x_2)$).
     Thus, the calculation of the power spectrum for the parking lot model
reduces to finding the distribution of relaxation times for adsorption, and
then weighting the spectrum for telegraph noise with this distribution.
    As shown above, the distribution function for the inter-car spaces is
exponential.  Assuming that the separations between a given car and its two
neighbors, $x_1$ and $x_2$, are uncorrelated, we obtain the following
distribution  of relaxation rates, $\nu=k_+(x_1+x_2)$:

\begin{equation} f(\nu)=\frac {\nu}{\omega_H^2}\exp\left(
-\frac{\nu}{\omega_H}\right)\end{equation}
Here

\begin{equation} \omega_H\equiv k_+\delta\end{equation}
is the characteristic frequency of the fast relaxation.

 One can now substitute the  distribution function for $\nu$ into the above
expression for the high--frequency fluctuation spectrum, Eq. (\ref{hi}).
The result of the numerical  integration of this formula over $\nu$ is in
excellent agreement with the simulations, see Figure 3.  We now have an
explanation for the hump in the spectrum at high frequency; it is caused by
a relatively broad distribution of relaxation times for the adsorption
process.  Further
examination of our result illuminates the connection between the PLM and other, more complicated systems
exhibiting glassy dynamics.
The above distribution of  relaxation rates does not just distort the
Lorentzian form near the characteristic frequency, $\omega_H\equiv
k_+\delta$, but rather  affects  the spectrum at all  frequencies below
$\omega_H$. The asymptotic behavior  of the spectrum at $\omega \ll
\omega_H$ is

\begin{equation}S(\omega)\sim \log \omega \end{equation}
Hence, the spectrum never recovers the Lorentzian--like plateau regime at
low frequencies. The logarithmic behavior is reminiscent of the power--law
spectrum typical for glasses at mesoscopic frequencies (it corresponds to
$\beta$--relaxation, \cite{glass}).
Because the logarithmic behavior is a  result of the problem being
one--dimensional, one could expect real power--law behavior in the
spectrum in higher dimensions. A simple  assumption that the adsorption
probability is proportional to the free volume associated with the newly
created void, combined with the natural exponential distribution for the
free volume, would result in the $1/\omega$ spectrum at mesoscopic frequencies: 
\begin{equation}S_H(\omega)\sim \int_0^\infty\frac{exp(-\nu/\nu_0)d\nu}{\omega^2+\nu^2}\sim 1/\omega\ \ ,\  \omega\ll\nu_0 \end{equation}

 Note that in the
one--dimensional case discussed above, the free volume is a sum of two
presumably independent variables, $x_1$ and $x_2$, each of which has an
exponential distribution. As a result, the distribution function for the
free volume vanishes near zero, i.e. the probability of finding a long--living
excited state is strongly suppressed in the one--dimensional case compared
to the higher dimensions.

\section{The Slow Dynamics of the System}

In the previous section we have  discussed the fluctuations of the density
on short time scales, over which the system remains ergodically broken. Now
we proceed to a discussion of the low--frequency part of the fluctuation
spectrum  associated with the change of the slow parameter $Z$. This change
is caused by collective events; the rearrangement of a state corresponding
to a given $Z$ is dominated by a two-car process. In one process,
responsible for decreasing the "ground state" $Z$ by $1$,  two adjacent
cars leave and a single one (a "bad" parker) comes in their stead, hogging
the space (see Figure 5a); the opposite process results in adding an extra
car to the lot: a car exits and leaves a large space--big enough for two
cars, provided that the new cars are "good" parkers (figure 5b).  How do
these collective modes affect the power spectrum?
     We  now calculate the rates of the above two--car processes. The rate
of the "$-1$" process has three contributions.  First, a  car must leave,
and the corresponding "trial rate" is just  $k_-$ per particle. Then,  an
adjacent car must leave before the hole left by the first car fills, which
gives a "waiting" factor
$2 k_-\int_0^\infty \int_0^\infty exp(-\nu t) dt f(\nu)d\nu$.
Finally, the  big hole must be blocked by a "bad" parker (up to a
correction of order of  $\delta$, the probability of this is unity). Thus
the overall rate is

\begin{equation}\nu_{-1}=\frac{2 k_-^2}{k_+\delta} \end{equation}

 The opposite, "$+1$"-- process, has  the same trial frequency, $k_-$ per
car. The void left by the car (its length is $x_1+x_2+1$) must be large
enough for two cars, which gives the factor

\begin{equation}\frac
{1+\Delta}{\delta^2}\exp\left[-\frac{1+\Delta}{\delta}\right] d \Delta
\end{equation}
Note that the first incoming car must park with precision
$\Delta=x_1+x_2-1$ in order to leave enough space for  the second car. The
probability of this happening contributes a factor of $2\Delta/(1+\Delta)$
to the overall "$+1$"--rate:

\begin{equation}\nu_{+1}= k_-\int_0^\infty \frac{2\Delta}
{\delta^2}\exp\left[-\frac{1+\Delta}{\delta}\right] d \Delta=2
k_-\exp(-1/\delta) \end{equation}
The density changes with time according to the following Equation:
\begin{equation}\label{mast}
\dot\rho=(\nu_{+1}-\nu_{-1})\rho+\eta(t)\end{equation}
Here $\eta$ is the noise originating from the fact that the density changes
by discrete one--particle steps. $\langle\eta(t)\rangle=0$, because the
average evolution is given by the interplay of the  "$+1$" and "$-1$"
kinetic terms in the above equation. Since adding an extra car or
removing one at the  moment of time $t_0$ corresponds to $\dot N=\pm
\delta(t-t_0)$, and since there is no obvious mechanism for the
correlations between such processes,

\begin{equation}\langle\eta(t)\eta(t')\rangle=\frac{\nu_{-1}+\nu_{+1}}{L}\rho
\delta(t-t')\end{equation}

Using Eq. (\ref{mast}), we find that the equilibrium density is determined
by the condition:

\begin{equation}\frac{k_-}{k_+}=\delta_{eq} \exp(-1/\delta_{eq})\end{equation}
where $\delta_{eq}=\rho_{eq}^{-1}-1$. This result coincides with the mean field one, Eq. (\ref{eq}). We note that the mean field
approach ignores the adsorption-desorption correlations; this would be  a
reasonable assumption for the model with  strong diffusion of the adsorbed
cars.  Since the diffusion cannot shift the equilibrium properties, it is not surprising that the mean field approach gives the correct value of
$\rho_{eq}$. As to the description of the PLM kinetics, the rates
$\nu_{-1}$ and $\nu_{+1}$ in Eq(\ref{mast}) differ by an exponentially small  factor,
$2\exp(-1/\delta_{eq})$ from their mean--field analogs. The above equation
describes only the slow evolution of the jammed state toward equilibrium.
Thus, the overall  density relaxation curve of the "cold" system (at
$k>k_c$) consists of the classical fast regime, Eq. (\ref{jam}),  resulting
in a jamming at $\rho=\rho_c\simeq 0.75$,  and the desorption--promoted
final stage, discussed here. Such a combination of the two theoretical
results is in agreement with the simulation data, as is shown in Figure 2.
The same figure shows that the mean field curve does not capture the two
stage nature of the relaxation dynamics and is inadequate for the
description of the cold system ($k=10^4$). However,  the mean field may  be
used for the description of the single--stage relaxation of the "hot"
system.

By expanding Eq. (\ref{mast}) near the equilibrium density, one can
determine the relaxation frequency of the system, $\omega_{L}$ and the
spectrum of the  low--frequency fluctuations:

\begin{equation}\omega_L=\frac{2k_+}{\delta_{eq}}\exp(-2/\delta_{eq})
\end{equation}
\begin{equation}S_L(\omega)=\frac{2k_- \rho_{eq}}{\pi
L}\frac{\exp(-1/\delta_{eq})}{\left(\omega^2+\omega_L^2\right)}\end{equation}

By combining this expression with the earlier result for the
high--frequency fluctuations, one obtains the following analytic form  for
the entire power spectrum of the PLM:

\begin{equation}S(\omega)\equiv S_H(\omega)+S_L(\omega)=\frac{k_-
\rho_{eq}}{\pi
L}\left(\int_0^\infty\frac{\exp(-\nu/\omega_H)}{\omega^2+\nu^2}\frac {\nu
d\nu}{\omega_H^2}+\frac{2\exp(-1/\delta_{eq})}{\omega^2+\omega_L^2}\right)
\end{equation}

This result agrees amazingly well with the simulation data as shown  in
Figure 3. Figure 6 shows how the characteristic frequencies $\omega_H$ and
$\omega_L$ depend on the control parameter $k\equiv k_+/k_-$ (or,
equivalently on $\rho_{eq}$); the theoretical calculation is again in good
agreement with simulation.

\section{Conclusions}

We have presented a theoretical discussion of the Parking Lot Model, which
appears to be a very simple glassy system, perhaps the simplest. We have
identified two time scales in the problem: one associated with a simple
relaxation of voids and the other corresponding to the collective
(two--particle) processes responsible for the rearrangement of the "ground
state" (the state that the system can reach by an instant filling of all
currently available voids).
In the limit of weak desorption, corresponding to a large difference
between the two times, the relaxation of the system toward its equilibrium
density is a two--stage process: first, it reaches the universal jamming
density $\rho=\rho_c\simeq 0.75$ as if there were no desorption at all,
then it slowly relaxes to $\rho_{eq}(k)$ via collective rearrangements.
This two-stage relaxation feature disappears in the regime of strong
desorption ($k<k_c$), and we identify the crossover point $k_c$ with
$\rho_{eq}(k)=\rho_c$.

At $k>k_c$, on times shorter than the  longest characteristic scale
($\omega_L^{-1}$) the evolution of the system is  non--ergodic; this regime
is analogous to a supercooled liquid. The system evolves by jumping between
the metastable states corresponding to different values of the parameter
$Z$. Note that the lifetime of these states does not grow with the system
size, but rather decreases. Since the probability of the rearrangement of
the ground state is $\nu_{+1}+\nu_{-1}$ per particle per unit time, its
lifetime is inversely proportional to the number of cars $N\simeq Z$:
$\tau_Z^{-1}=Z(\nu_{+1}+\nu_{-1})$. In this sense,  the free energy
landscape of PLM is similar to that of structural glasses.
Another similarity between the PLM and  glassy systems is the possibility
of accelerating relaxation by means of  annealing.

The existence of two characteristic time scales is responsible for the
intriguing form of the fluctuation spectrum of the reversible parking
problem. The slow fluctuations are described by a single Lorentzian with
the relaxation frequency $\omega_L$ associated with two--particle
rearrangements. The fast dynamics is a superposition of many
single--particle adsorption--desorption modes.  In a  sense, the non--Lorentzian form of the high--frquency part of the  power spectrum is a
reflection of the deviations of the local density from its average value,
i.e. it is a signature of long--living disorder. In this form, our
observation may be relevant for understanding the non--trivial behavior of
the relaxation spectrum of other glassy systems on mesoscopic frequencies.
The distribution of the
relaxation rates of the single--particle excitations  results not only in the distortion of the Lorentzian in
the vicinity of the characteristic corner frequency $\omega_H$, but also in
an interesting logarithmic behavior of the power spectrum at lower
frequencies ($S\sim \log \omega$). We have suggested that this feature is
reminiscent of the power law spectrum corresponding to $\beta$--relaxation
in classical glasses and that such a power--law behavior could be
reproducible in PLM at higher dimensions.

\bigskip

{\bf Acknowledgments}

\medskip

We thank H. M.  Jaeger, S. R.  Nagel, L. P. Kadanoff, S. Coppersmith, T. A.
Witten, J. Cederberg, B. Cipra, and M. Richey for inspiring and valuable
discussions. The work is supported in part through the MRSEC program by the
National Science Foundation through grants NSF DMR 9400379, NSF DMR 9415605 and NSF DMR 9528957.   We express our gratitude to the Pew Midstates
Science and Mathematics Consortium for funds provided through their short term consultation program and to St. Olaf College for support through its sabbatical leave program.

\newpage\

\newpage

\centerline {\bf Figure Captions}

\begin{description}

\item{Figure 1.} The parking lot model.
\item{Figure 2.} Simulation data (circles) and theoretical results  for
density evolution in the "cold" ($k=10^4$) amd "hot"($k=10$) regimes. Note
that the mean field (dashed line) is quite adequate for the description of
the "hot" system but it fails to describe the slow dynamics at the
"supercooled" regime for $k=10^4$. The solid line is the combination of the
irreversible parking curve, Eq (1), describing the fast stage of the
relaxation (for $k=10^4$),  and our result, beginning at $t=100$, for the later slow dynamics.

\item{Figure 3.} The power spectrum of density fluctuations near the
equilibrium for $k=10^3$ and $k=10^4$. Circles and solid lines represent
the simulation data and our analytic results, respectively.
\item{Figure 4.} "Kinetic Phase Diagram" of the system. The solid line
corresponds to the reversible equilibrium line $\rho_{eq}(k)$, the dashed
line shows the jamming density above which the ergodicity is partially
broken.
\item{Figure 5.} The slow collective rearrangements corresponding to (a) adding
 and (b) subtracting  one car.
\item{Figure 6.} The low  and high characteristic frequencies, $\omega_L$
and $\omega_H$, as functions of the equilibrium density. The dashed and
solid lines represent the theoretical results, the squares and the circles
are the simulation data for $\omega_L$ and $\omega_H$, respectively.

\end{description}

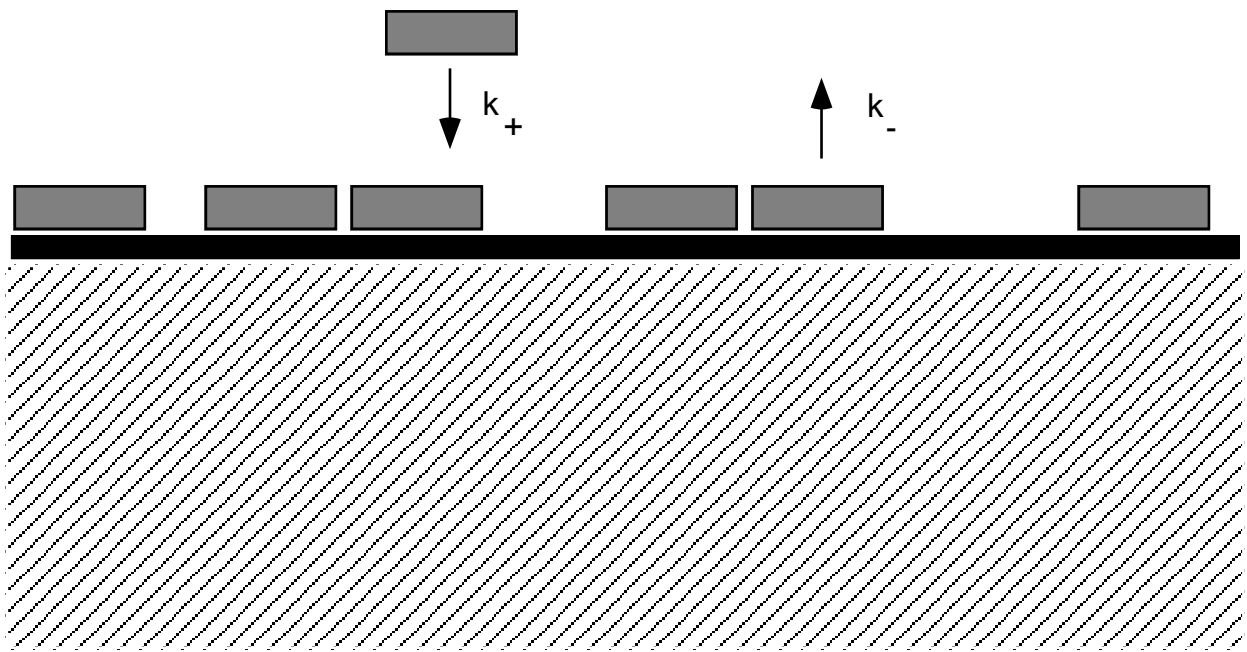
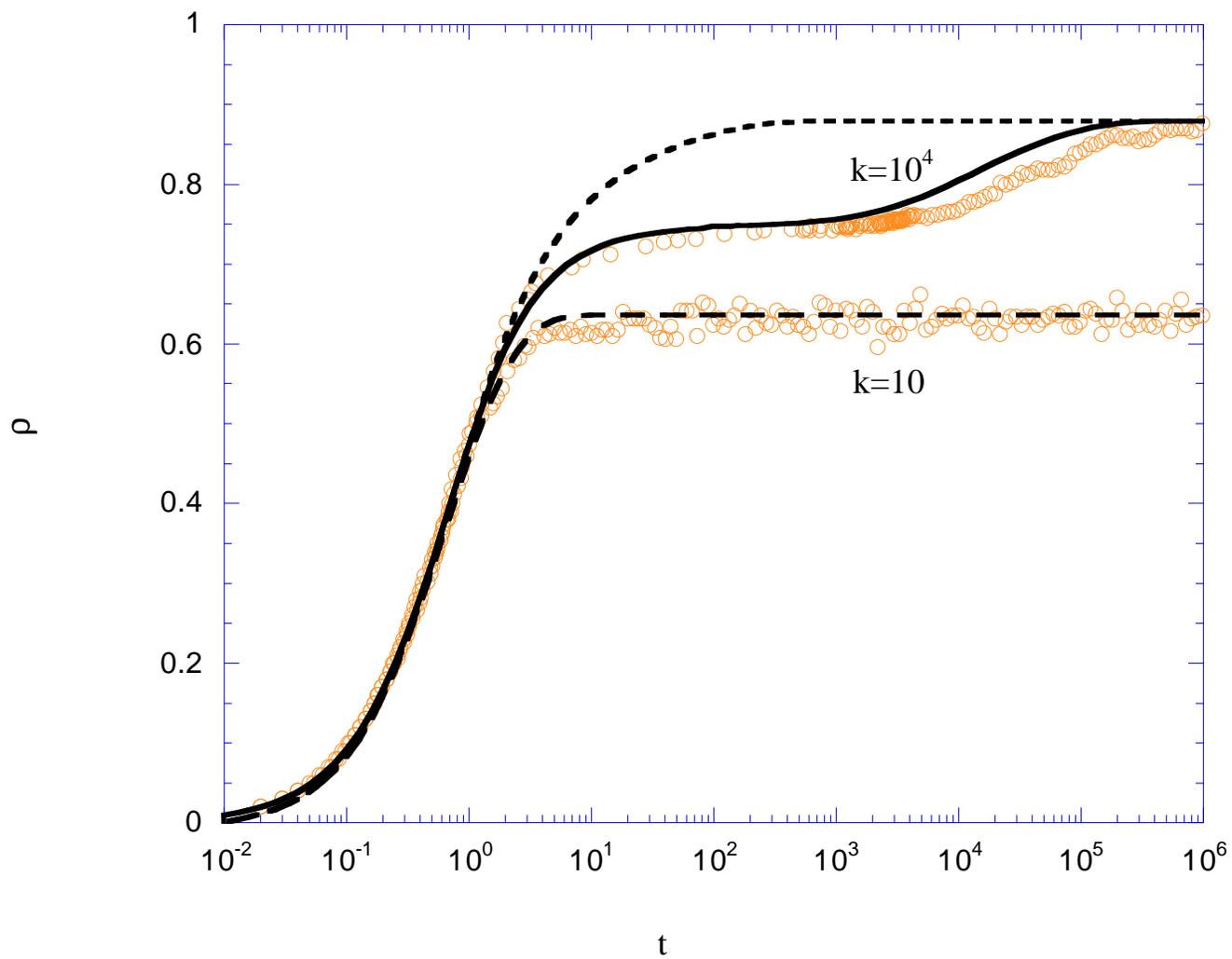
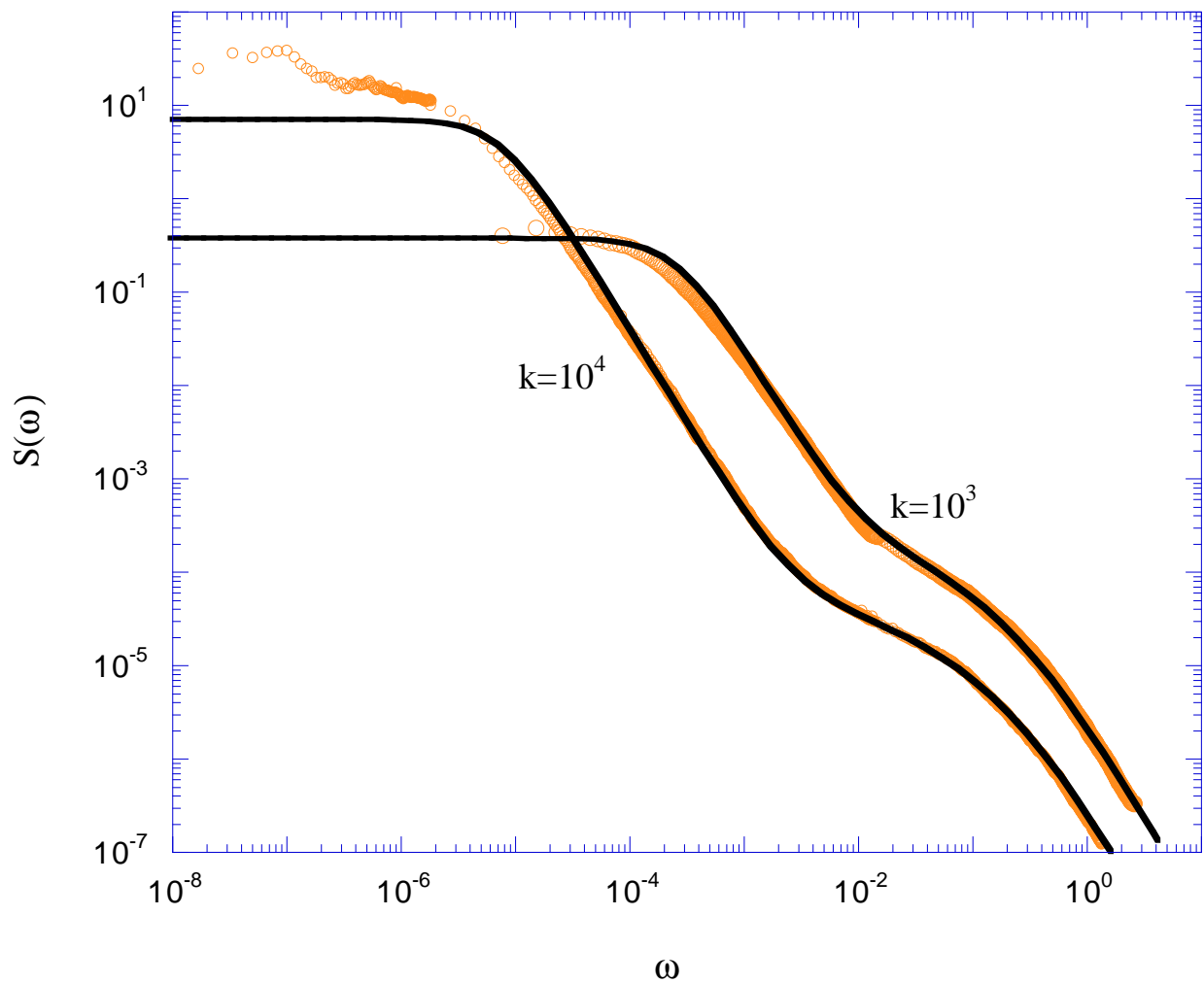
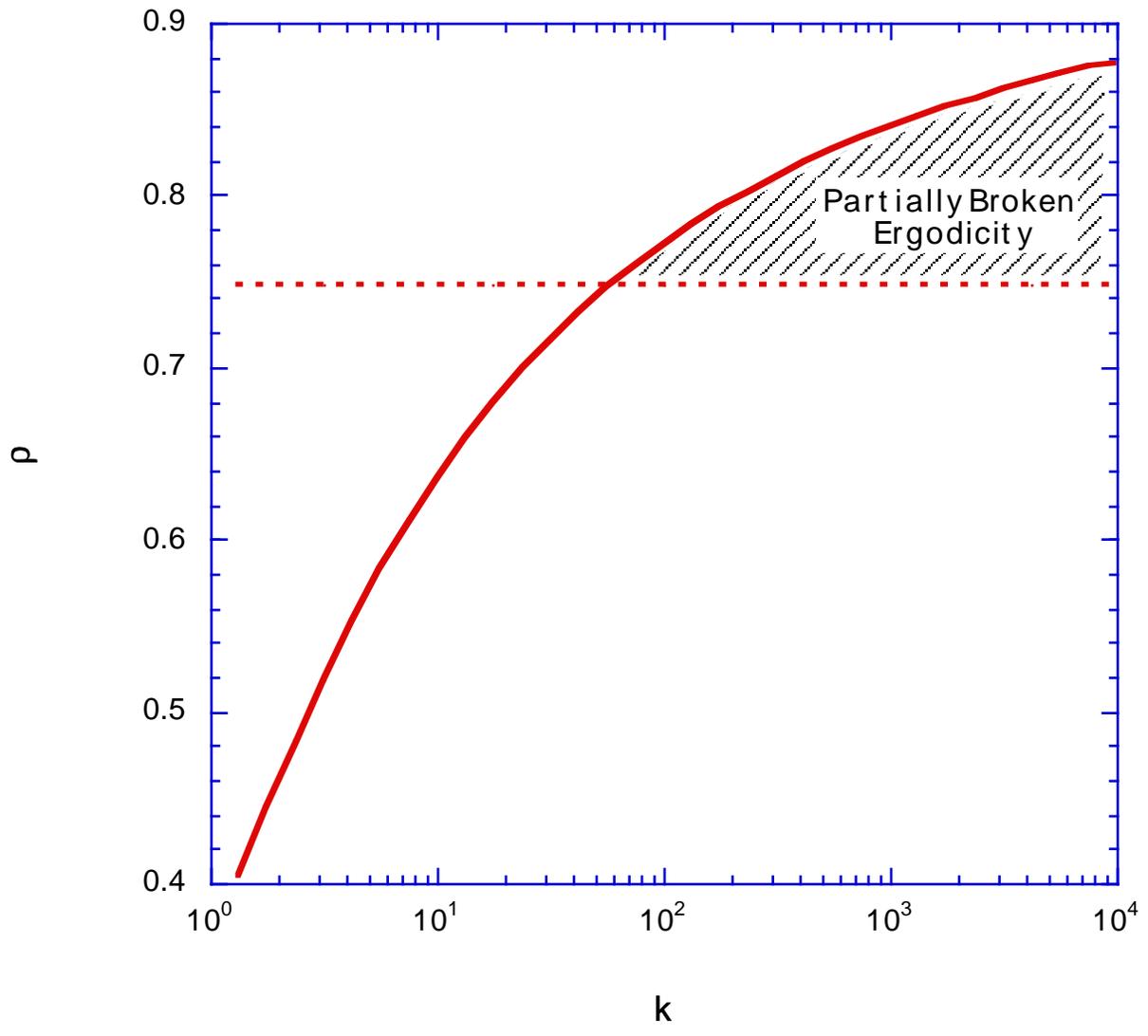
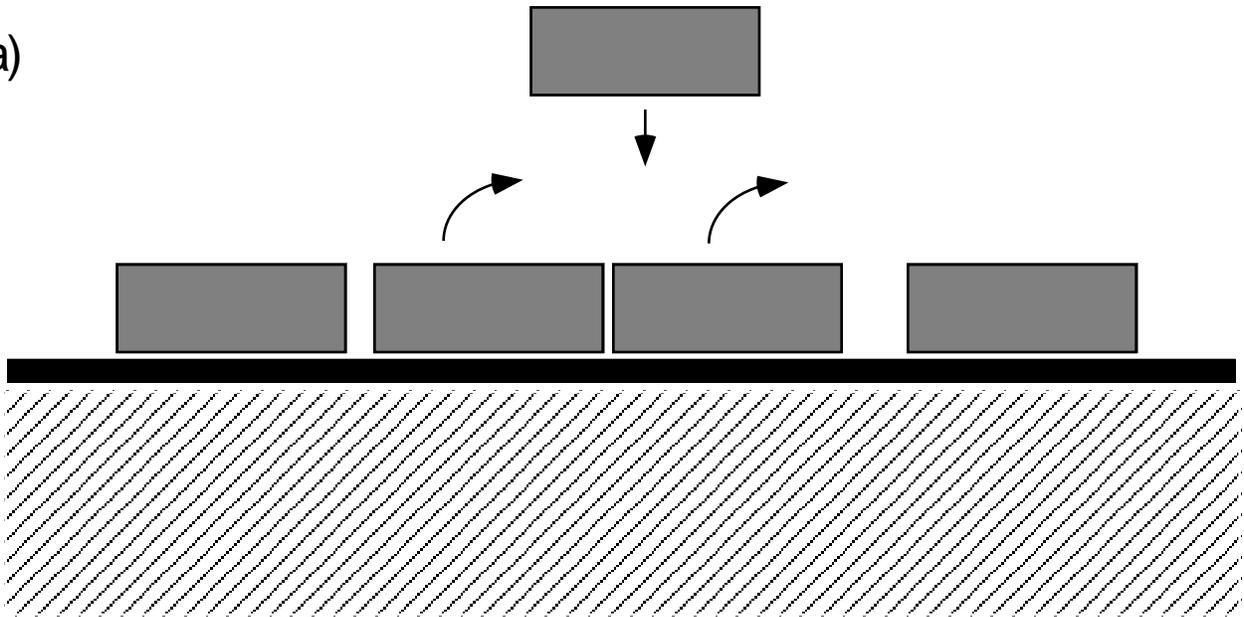
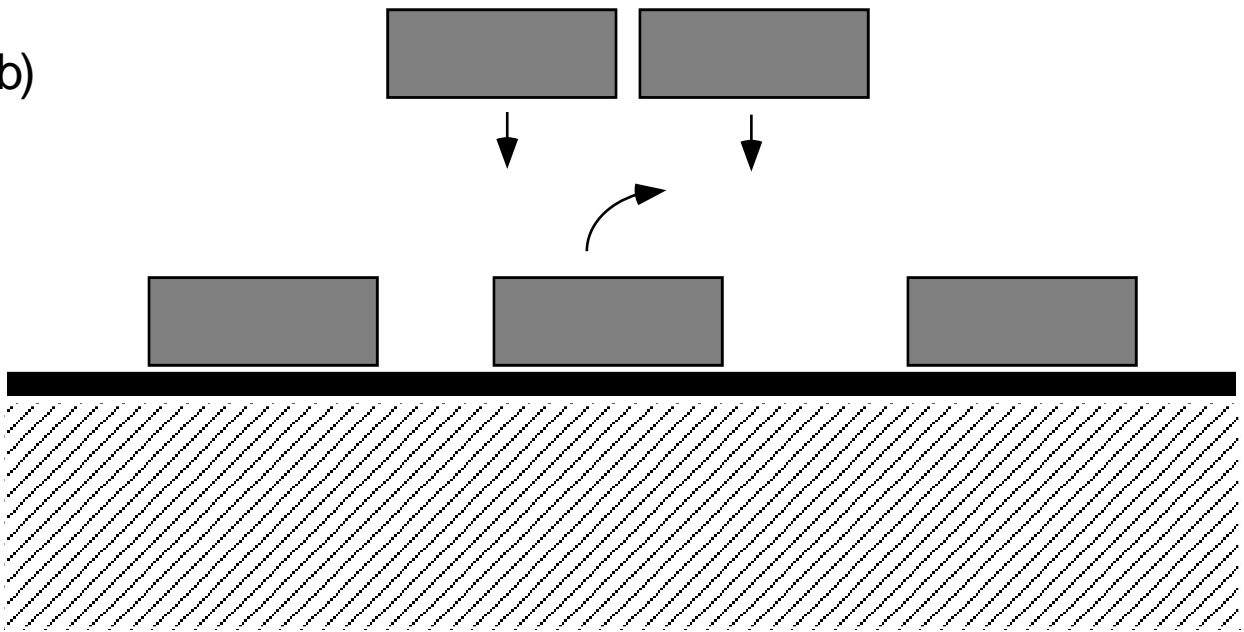
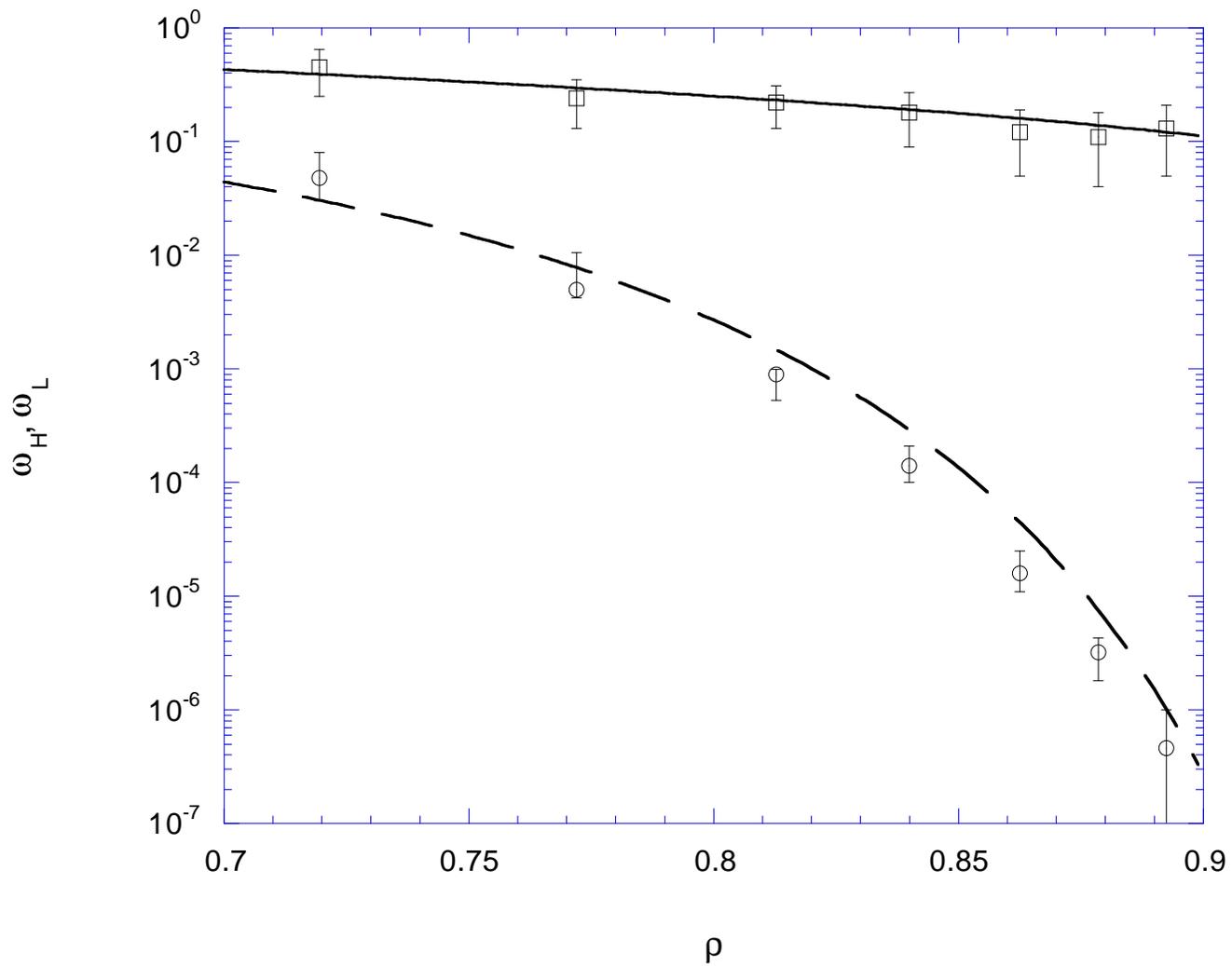
\end{document}